\documentclass[conference]{IEEEtran}

\usepackage[utf8]{inputenc} 
\usepackage[T1]{fontenc}    
\usepackage{url}            
\usepackage{booktabs}       
\usepackage{amsfonts}       
\usepackage{nicefrac}       
\usepackage{microtype}      
\usepackage{pgfplots}

\usepackage[caption=false,font=footnotesize]{subfig}

\usepackage{amsthm}

\usepackage{tikz}
\usetikzlibrary{calc}
\usepackage[short]{optidef}

\usepackage{algorithm}
\usepackage[noend]{algpseudocode}
\makeatletter
\def\BState{\State\hskip-\ALG@thistlm}
\makeatother
\usepackage[font=small,labelfont=bf]{caption}
\usepackage{mathtools}

\DeclarePairedDelimiter\floor{\lfloor}{\rfloor}

\newcommand{\Aa}{{N_x}}
\newcommand{\AB}{{N_z}}
\newcommand{\Bb}{{N_y}}
\newcommand{\LY}{{L}}      
\newcommand{\ND}{{N}}      
\newcommand{\DM}{{K}}      
\newcommand{\Input}[1]{ \textbf{Input}:#1}

\usepackage{tabularx}
\usepackage{makeidx}
\makeindex 
\makeatletter
\newcommand{\multiline}[1]{%
	\begin{tabularx}{\dimexpr\linewidth-\ALG@thistlm}[t]{@{}X@{}}
		#1
	\end{tabularx}
}
\makeatother

\title{Hierarchical coded matrix multiplication}
	\author{\IEEEauthorblockN{Shahrzad Kiani, Nuwan Ferdinand, Stark C. Draper}\\
		\IEEEauthorblockA{Department of Electrical and Computer  Engineering, University of Toronto, Toronto, ON, Canada}
		\IEEEauthorblockA{Email:shahrzad.kianidehkordi@mail.utoronto.ca\{nuwan.ferdinand and stark.draper\}@utoronto.ca }}

\begin{document}

\maketitle

\begin{abstract}

Slow working nodes, known as \emph{stragglers}, can greatly reduce the
speed of distributed computation. Coded matrix multiplication is a
recently introduced technique that enables straggler-resistant
distributed multiplication of large matrices. A key property is that
the finishing time depends only on the work completed by a set of the
fastest workers, while the work done by the slowest workers is ignored
completely. This paper is motivated by the observation that in
real-world commercial cloud computing systems such as Amazon's
Elastic Compute Cloud (EC2) the distinction between fast and slow nodes is
often a soft one. Thus, if we could also exploit the work completed by
stragglers we may realize substantial performance gains. To realize
such gains, in this paper we use the idea of hierarchical coding
(Ferdinand and Draper, {\em IEEE Int. Symp. Inf. Theory}, 2018).  We
decompose the overall matrix multiplication task into a hierarchy of
heterogeneously sized subtasks. The duty to complete each subtask is
shared amongst all workers and each subtask is (generally) of a
different complexity. The motivation for the hierarchical
decomposition is the recognition that more workers will finish the
first subtask than the second (or third, forth, etc.). Connecting to
error correction coding, earlier subtasks can therefore be designed to
be of a higher rate than later subtasks. Through this hierarchical
design our scheme exploits the work completed by stragglers, rather
than ignoring it, even if that amount is much less than that completed
by the fastest workers. We numerically show that our method realizes a
$60\%$ improvement in the expected finishing time for a widely studied
statistical model of the speed of computation and, on Amazon EC2, the
gain is $35\%$.
\end{abstract}

\section{Introduction}
Many data intensive problems cannot be solved in a single computer due to limited processing power and storage. Distributed computation is necessary. In an idealized distributed setting one would expect highly parallelizable workloads to realize an acceleration proportional to the number of nodes. However, in cloud-based systems, slow workers, known as \emph{stragglers}, are a bottleneck that can prevent the realization of faster compute times~\cite{Dean:2012}. Recent studies show that for workloads that are linear algebraic in nature, the effect of stragglers can be minimized through the use of error correction codes~\cite{Lee:TIT17, YuEtAl:2017, Alex2017:GC, Lee:MATRIXISIT17,  Dutta:ISIT17, DuttaEtAl:2018}. The underlying idea is to introduce redundant computations (additional workers are needed) such that the completion of any fixed-cardinality subset of jobs suffices to realize the desired solution. 

A trivial approach to introducing redundancy is through replication. But, if a task can be linearly decomposed, the opportunity arises to introduce redundancy through the use of error correction codes such as maximum distance separable (MDS) codes. In~\cite{Lee:TIT17}, it was shown that MDS codes can be leveraged to design straggler-resistant methods of vector-matrix multiplication. The ideas were extended to matrix-matrix multiplication based on the product codes in~\cite{Lee:MATRIXISIT17}. In~\cite{YuEtAl:2017}, coded computation based on polynomial interpolation was introduced. Such polynomial codes outperform product codes in terms of their \emph{recovery threshold}. The recovery threshold is the number of workers that must complete their tasks to realize the computation. The recovery threshold was further improved in~\cite{DuttaEtAl:2018} through a different approach to matrix multiplication, called MATDOT codes.

A drawback of all these methods is that they rely on the work completed by a set of the fastest workers, ignoring completely work completed by the slower workers. Effectively, in these methods, stragglers are modeled as extremely slow nodes that complete no work. In the terminology of error correction coding, they are modeled as erasures. However, in cloud base systems such as the Amazon Elastic Compute Cloud (Amazon EC2), we observe \emph{partial} stragglers. Partial stragglers are slower, only able to complete partial tasks by the time at which the faster workers have completed their entire tasks. That said, the amount of work stragglers can complete may be non-negligible. Thus, it can be wasteful to ignore.

The concept of exploiting stragglers has been studied in~\cite{Alex2017:GC,  EXPLOIT:ISIT18, RATELESS:2018, HIER:ISIT18}. All these papers consider a distributed system with a central node, called the master, and multiple worker nodes. The partial-straggler scenario was first considered in~\cite{ Alex2017:GC} such that each worker was assigned two groups of subtasks: {\em naive} and {\em coded} subtasks. Non-straggler workers processed both naive and coded subtasks, while partial stragglers only completed  naive tasks. In~\cite{ EXPLOIT:ISIT18, RATELESS:2018}, the vector matrix multiplication was first broken into computationally homogeneous subtasks; each subtask was then encoded using a specific code: an MDS code in~\cite{ EXPLOIT:ISIT18} and a rateless fountain code in~\cite{ RATELESS:2018}. The central idea in exploiting stragglers is to assign each worker multiple small subtasks rather than a single large task. The master is able to complete the job by utilizing the subcomputations completed by all workers, stragglers simply contribute less. The coding method used in~\cite{ EXPLOIT:ISIT18, RATELESS:2018} enables a trade-off between the size of each subtask and the recovery threshold. This means that while the size of each subtask is $1/\LY$ of the size of a single task, the master requires $\LY$ times more subtasks to be completed. The increase in the recovery threshold increases the encoding and decoding complexity. To reduce the complexity overhead of encoding and decoding while preserving the goal of exploiting stragglers, \cite{HIER:ISIT18} introduced the concept of {\em hierarchical coding} by leveraging the \emph{sequential} computing nature of workers. In~\cite{HIER:ISIT18}, hierarchical coding was applied to the vector matrix multiplication using MDS codes.

In this paper, we apply the idea of hierarchical coding to matrix multiplication. Before introducing the general framework of hierarchical coded matrix multiplication, it helps to recapitulate the intuition behind~\cite{HIER:ISIT18}. In hierarchical coding, the total computation required of each worker is partitioned into \emph{layers} of sub-computations. Workers compute layers sequentially. Due to the sequential processing, each layer has a different finishing time, i.e, a worker will start to work on the second layer after it finishes the first layer. Therefore, the finishing time of the first layer is lower than that of the second layer. Each layer is encoded using a different code with a distinct rate such that all layers have the same expected finishing time. 

To extend the above idea to matrix multiplication we establish an equivalence between task allocation in distributed matrix multiplication and a geometric problem in which partitioning the matrix to be computed is visualized as partitioning a rectangle into tiles. Larger tiles correspond to higher rate  subtasks that each worker tackles first. The choice of the area of each tile is set according to the statistics of the computing system.  The overall packing of tiles into the matrix is posed as an optimization problem. We encode each layer using different polynomial\footnote{Although we can also apply hierarchical coding to other codes, such as MATDOT codes~\cite{DuttaEtAl:2018}, due to space constraints herein we focus on introducing hierarchical coded matrix multiplication in the context of polynomial codes.} codes with distinct recovery thresholds.  In our system it does not matter which workers complete each subtask, but we do need to know how many workers are expected to complete each subtask. Our method realizes a $60\%$ reduction in expected finishing time when compared to the baseline scheme of non-hierarchical polynomial coding~\cite{YuEtAl:2017}.

\section{Problem statement and system model}
\label{model}
In this section, we first provide an example to illustrate the
intuition behind our scheme. We then develop the general hierarchical coded computation scheme in detail.
\subsection{Motivating example}
\label{exp}

Consider the task of multiplying two matrices, $A \in \mathbb{R}^{\Aa \times \AB}$ and $B \in \mathbb{R}^{\AB\times \Bb}$. The computation task $AB \in \mathbb{R}^{\Aa\times \Bb}$ requires $\Aa\AB\Bb$ multiply-and-accumulate operations. Let us suppose that we want to parallelize this task among $\ND \geq 8$ worker nodes, by providing a number of smaller subtasks to each. As the subtasks will be smaller, individual nodes will require less time to finish their subtasks as compared to computing the entire $AB$ product; parallelization will thereby be exploited. 

In the following three scenarios we first detail the usage of polynomial codes~\cite{YuEtAl:2017} to solve this problem. We then present our proposed hierarchical coded matrix multiplication. In Scenario 3, we compare our scheme with the state-of-the-art exploit-straggler scenario~\cite{EXPLOIT:ISIT18}, referred as sum-rate codes. We comment that, the original idea of sum-rate coding~\cite{EXPLOIT:ISIT18} was introduced through the vector-matrix multiplication problem using MDS codes~\cite{Lee:TIT17} and was extended to the matrix multiplication problem using product codes~\cite{Lee:MATRIXISIT17}. However, here we keep on using polynomial codes as the baseline, to which sum-rate coding can be easily applied. In all cases, we allocated $\frac{\Aa\AB\Bb}{4}$ multiply-and-accumulate operations in total to each worker. This makes each worker's computational load equal for a fair comparison.

\textbf{Scenario 1 (polynomial codes):} In the first scenario, we use polynomial codes~\cite{YuEtAl:2017}. We decompose the rows of $A$ and the columns of $B$ into two equally sized groups, $A^T=[{A}_1^T, {A}_2^T]$ and $B=[{B}_1, {B}_2]$, $A_i \in \mathbb{R}^{\frac{\Aa}{2} \times \AB}$ and $B_i \in \mathbb{R}^{\AB \times \frac{Bb}{2}}$ where $i \in \{ 1,2\}$. We then assign the multiplication of $({A}_1 + {A}_2 n) \in  \mathbb{R}^{\frac{\Aa}{2} \times \AB}$ and $({B}_1 + {B}_2 n^2) \in \mathbb{R}^{\AB \times \frac{\Bb}{2}}$ to the $n$-th worker $n \in [\ND]$, where $[\ND]=\{1,\ldots,\ND\}$ is the index set of cardinality $\ND$. I.e., the $n-$th worker computes $({A}_1 + {A}_2 n)({B}_1 + {B}_2 n^2)$ requiring $\frac{\Aa\AB\Bb}{4}$ operations. It can be shown that the outputs of any $\DM_{\text{poly}}=4$ of the $\ND$ workers is sufficient to recover the desired $AB$ matrix computation via polynomial interpolation~\cite{YuEtAl:2017}. We refer to $\DM_{\text{poly}}/\ND$ as the rate of polynomial code.


\textbf{Scenario 2 (hierarchical polynomial codes):} In the second scenario, we introduce hierarchical coded matrix multiplication and show it outperforms the  polynomial coded approach of Scenario $1$. In our setting each worker is provided $\LY = 4$ sequentially-ordered subtasks, each of which involves $\frac{\Aa\AB\Bb}{16}$ multiply-and-accumulate operations such that the total computation load matches that of Scenario 1. Upon completion of each subtask, the result is sent to a master node to derive the final result $AB$. Importantly, in our proposed scheme, not all workers need complete all their subtasks for the master to be able to compute the product $AB$. Instead, we allow workers to finish different number of subtasks. We term the choice of partitioning the \emph{profile} of subtasks. The profile is something that we will optimize based on the statistics of the distribution of computational speeds of the workers. Such optimization will be discussed further in a later section. For illustrative reasons, in this example we assume the $(8, 4,3,1)$ profile. That is, only $\DM_1=8$ workers need to complete their first subtask, only $\DM_2=4$ workers need to finish their first and second subtasks, and to complete layer $3$ and $4$, only $\DM_3=3$ and $\DM_4=1$ workers are required to complete their subtasks in those layers. 

To generate the subtasks, we first partition the product matrix $AB$ into $\LY=4$ rectangular \emph{task} tiles each of which we think of as a layer of computation. Each element of each task tile is an inner product of a row of $A$ and a column of $B$. We partition the task tile corresponding to the $l-$th layer, $l \in [\LY] = [4]$ into $\DM_l$ equally sized \emph{information} tiles each of which contains roughly $\frac{\Aa\Bb}{16}$ inner products. Hence, the area of task tile corresponding to $l$th layer is equal to $a_l=\frac{\DM_l \Aa\Bb}{16}$.

In~\ref{eq.basicEx}, we depict one possible partitioning into task tiles. For example, the task tile corresponding to the first layer consists of $\frac{8\Aa\Bb}{16}$ inner products involving all rows of $A$ and the first half of the columns of $B$. The task tile is further partitioned into $\DM_1=8$ equally sized information tiles.  These information tiles $A_{1,i}B_{1,j}$, $i \in [4]$ and $j \in [2]$ are the product of \emph{data chunks} $A_{1,i} \in \mathbb{R}^{\frac{\Aa}{4} \times \AB}$ and $B_{1,j} \in \mathbb{R}^{\AB \times \frac{\Bb}{4}}$. The other layers $l \in \{2, 3, 4\}$ similarly partition their task tiles into information tiles. In this example the number of information tiles for these layers are, respectively, of 4,3, and 1.

\begin{align}\label{eq.basicEx} 
AB=
 \left[ \begin{array}{cccc}
   A_{1,1} B_{1,1} & A_{1,1} B_{1,2} & A_{2,1} B_{2,3} & A_{3,1} B_{3,4}\\
   A_{1,2} B_{1,1} & A_{1,2} B_{1,2} & A_{2,2} B_{2,3} & A_{3,2} B_{3,4}\\
   A_{1,3} B_{1,1} & A_{1,3} B_{1,2} & A_{2,3} B_{2,3} & A_{3,3} B_{3,4}\\
   A_{1,4} B_{1,1} & A_{1,4} B_{1,2} & A_{2,4} B_{2,3} & A_{4,4} B_{4,4}
\end{array}\right]
\begin{tikzpicture}[remember picture,overlay]
\draw [dashed] ($(-6.75,-0.7)$) to ($(-3.7,-0.7)$) to ($(-3.7,0.9)$) to ($(-6.75,0.9)$) to ($(-6.75,-0.7)$);
\draw [dashed] ($(-3.5,-0.7)$) to ($(-2.05,-0.7)$) to ($(-2.05,0.9)$) to ($(-3.5,0.9)$) to ($(-3.5,-0.7)$);
\draw [dashed] ($(-1.8,-0.27)$) to ($(-0.4,-0.27)$) to ($(-0.4,0.9)$) to ($(-1.8,0.9)$) to ($(-1.8,-0.27)$);
\draw [dashed] ($(-1.8,-0.7)$) to ($(-0.4,-0.7)$) to ($(-0.4,-0.32)$) to ($(-1.8,-0.32)$) to ($(-1.8,-0.7)$);
\end{tikzpicture}
\end{align}

The master encodes the data relevant to $l$th task tile by applying a pair of polynomial codes (separately) to the data chunks involved in that tile. Thus, we apply the polynomial-code-based-approach first described in~\cite{YuEtAl:2017} on a per-task-tile basis.  The polynomials thus formed are the \emph{encoded} data chunks $\hat{A}_l(x)$, and $\hat{B}_l(x)$,
$l \in [\LY]$. For the above example, $\hat{A}_1(x) = A_{1,1} + A_{1,2} x + A_{1,3} x^2 + A_{1,4} x^3$, $\hat{B}_1(x) = B_{1,1} + B_{1,2} x^4$, $\hat{A}_2(x) = A_{2,1} + A_{2,2} x + A_{2,3} x^2 + A_{2,4} x^3$, $\hat{B}_2(x) = B_{2,3}$, $\hat{A}_3(x) = A_{3,1} + A_{3,2} x + A_{3,3} x^2$, $\hat{B}_3(x) = B_{3,4}$, $\hat{A}_4(x) = A_{4,4} $, and $\hat{B}_4(x) = B_{4,4}$.


Worker $n \in [\ND]$ gets $\LY=4$ pairs of encoded data chunks $(\hat{A}_l(n), \hat{B}_l(n))$ for all $l \in [\LY]$. The job of each worker is to compute the \emph{encoded} products $\hat{A}_l(n)\hat{B}_l(n) \in \mathbb{R}^{\frac{\Aa}{4} \times \frac{\Bb}{4}}$, working through $l\in [\LY]$ sequentially (in order) from $1$ to $\LY$, transmitting each result to the master as it is completed.
 
In above example the master can recover the first task tile as long as it received $\DM_1=8$ encoded products $\hat{A}_1(n) \hat{B}_1(n)$ from any of the $\ND$ workers. Similarly it can recover the second task tile as long as it receives $\DM_2=4$ encoded products $\hat{A}_2(n) \hat{B}_2(n)$ from any $4$ of the $\ND$ workers, and so forth. The ability of the master to decode from any sufficiently-large subset is a property of polynomial codes. E.g., the polynomial $\hat{A}_1(x)\hat{B}_1(x)$ is a polynomial of degree $7$ and therefore it can be recovered via polynomial interpolation as long as at least $8$ distinct values are known (In our setting, the values correspond to the indices of the workers that respond.).

The advantage of our scheme follows from the different rate applied across the jobs, $8/\ND$, $4/\ND$, $2/\ND$, and $1/\ND$ in this example and $\frac{\DM_l}{\ND}$ in general. 


\textbf{Scenario 3 (sum-rate polynomial codes):} In sum-rate
polynomial codes we use the same partitioning as the $AB$ matrix
depicted in~\ref{eq.basicEx}, but the data chunks are used to generate
a single polynomial code (instead of four).  The recovery threshold
of this code is $\DM_{\text{S-poly}}=16$. Note that we
use the term \emph{sum-rate} for this approach (originally developed
in~\cite{EXPLOIT:ISIT18}) to highlight the fact that the rate of this code,
$\DM_{\text{S-poly}}/\ND$, is equal to the sum of the per-layer rates
used in the hierarchical code.  In the above example
$\DM_{\text{S-poly}}/\ND = \sum_l \DM_l/\ND$ and $16/\ND = 8/\ND +
4/\ND + 3/\ND + 1/\ND$.

In the sum-rate approach we reproduce 
the $16$ sub-computations into which $AB$ is divided in~(\ref{eq.basicEx}) by dividing each of the $A$ and $B$ matrices into four equally sized sub-matrices.  Respectively, these are $A^T=[{\tilde{A}}_1^T, {\tilde{A}}_2^T, {\tilde{A}}_3^T, {\tilde{A}}_4^T]$ and $B=[{\tilde{B}}_1, {\tilde{B}}_2, {\tilde{B}}_3, {\tilde{B}}_4]$, where $\tilde{A}_i \in \mathbb{R}^{\frac{\Aa}{4} \times \AB}$ and $\tilde{B}_i \in \mathbb{R}^{\AB \times \frac{Bb}{4}}$ for $i \in [4]$.  After  partitioning, the master encodes the $\tilde{A}_i$ and the $\tilde{B}_i$ separately using polynomial codes, to generate encoded sub-matrices $\hat{A}(x) = \tilde{A}_1 + \tilde{A}_2 x + \tilde{A}_3 x^2 + \tilde{A}_4 x^3$ and $\hat{B}(x) = \tilde{B}_1 + \tilde{B}_2 x^4 + \tilde{A}_3 x^8 + \tilde{A}_4 x^{12}$. Worker $n$ is then tasked to compute $4$ sequentially-ordered subtasks: $\{\hat{A}(4n+i)\hat{B}(4n+i) | i \in [4] \}$. Due to the use of polynomial codes, the completion of any $16$ subtasks enables the recovery of the $AB$ product. 

Sum-rate codes have a more flexible recovery rule than do hierarchical codes. While for sum-rate codes, the $AB$ product can be recovered from any $16$ completed subtasks, in hierarchical codes the completed subtasks must follow a specific profile $(8,4,3,1)$. However, the completion time statistics of hierarchical codes can be quite close to that of sum-rate codes if the profile of hierarchical codes is designed correctly.

From a decoding perspective, the hierarchical approach is much less
complex than the sum-rate approach. In sum-rate codes, the master
needs to deal with decoding $\Aa\Bb/16$ polynomials of degree $16$. On
the other hand, in hierarchical codes the master is required to decode
four sets of polynomials of (in the example) degrees $8,4,3$ and $1$,
each set consisting of $\Aa\Bb/16$ polynomials. Furthermore, when
hierarchical codes are employed decoding can be carried out either in
a serial manner or parallelized across layers. Parallel decoding is
not possible for sum-rate codes. In the numerical results of
Sec~\ref{sim}, we will be observe that even serial decoding of
hierarchical codes takes less time than decoding sum-rate codes. As
would be guessed, the parallel decoding time of hierarchical codes is
less than the decoding time of sum-rate codes. This is due to the fact
that in the decoding phase of hierarchical codes, in a worst-case
scenario, the master needs to deal with decoding a polynomial code of
rate $8/\ND$.  This is much less computationally intensive than the
decoding of the rate $16/\ND$ polynomial code used in sum-rate codes.

Note that while the $\tilde{A}_i$ in sum-rate codes is equal to the $A_{l,i}$ in scenario 2, for all $l \in [4]$ (and similarly $\tilde{B}_i= B_{l,i}$), the extension of hierarchical coding to the general matrix multiplication problem is not as straightforward as it is for sum-rate codes. As will be described in next section, to make the hierarchical generalization, we take advantage of a useful geometric visualization in terms of partitioning the matrix to be computed into tiles.

\subsection{Hierarchical coded matrix multiplication} \label{general}
We now present our general construction.  Our goal is to compute the matrix $AB$ where, as before, $A \in \mathbb R^{\Aa \times \AB}$ and $B \in \mathbb R^{\AB \times \Bb}$. Our system consists of a master and $\ND$ workers.

We start by partitioning $AB$ into $\LY$ layers of computation.  As detailed below, computations relevant to each of the $\LY$ layers are shared with all $\ND$ workers.  Workers start working on layer $1$ and progress sequentially layer-to-layer. Each layer has a geometric interpretation as a rectangular task tile of elements of the $AB$ matrix. The $l$th such tile is described by the set $\mathcal{S}_l = \{\mathcal{S}_{Al}, \mathcal{S}_{Bl}\}$ where $\mathcal{S}_{Al}$ is a subset of consecutive elements of $[\Aa]$ and $\mathcal{S}_{Bl}$ is a subset of consecutive elements of $[\Bb]$. The $l$th task tile consists of the set of inner products of the $i$th row of $A$ and the $j$th column of $B$ where $i \in \mathcal{S}_{Al}$ and $j \in \mathcal{S}_{Bl}$.
The task tile can be visualizes as a rectangle of dimensions
$|\mathcal{S}_{Al}| \times |\mathcal{S}_{Bl}|$. We use $a_l$ to
denote the (integer) area of the $l$th tile, i.e., $a_l =
|\mathcal{S}_{Al}|\,|\mathcal{S}_{Bl}|$. To ensure that the tiles partition the entire $AB$ matrix these sets must satisfy $\cup_{l \in [\LY]} \{(i,j) \in \mathcal{S}_{Al} \times\mathcal{S}_{Bl}\} = [\Aa]\times [\Bb]$ and thus, $\sum_{l \in [\LY]} |\mathcal{S}_{Al}||\mathcal{S}_{Bl}| = \sum_{l \in \LY} a_l = \Aa\Bb$\footnote{We comment that task tiles can be allowed to overlap; that simply would mean certain elements of the matrix $AB$ would be computed in more than one of the task-layers.}. To denote the $i$th element of $\mathcal{S}_{Al}$ we write $\mathcal{S}_{Al,i}$, which is a row-index into the $A$ matrix. Similarly, $\mathcal{S}_{Bl,j}$ is a column-index into the $B$ matrix.

To apply error-correction coding to the computation of $AB$ we divide each task tile (corresponding to one layer of computation) into equally sized information tiles. To understand what we mean by information tile, we start by subdividing the inputs required to compute each task tile. The data required to compute the $l$th task tile (computation layer $l$) consists of the rows of $A$ and the columns of $B$, respectively, indexed by $\mathcal{S}_{Al}$ and $\mathcal{S}_{Bl}$. We use $A_{\mathcal{S}_{Al}}$ ($B_{\mathcal{S}_{Bl}}$) to denote the rows (columns) of $A$ ($B$) indexed by $\mathcal{S}_{Al}$ ($\mathcal{S}_{Bl}$). We next partition $A_{\mathcal{S}_{Al}}$ ($B_{\mathcal{S}_{Bl}}$) into $M_{xl}$ ($M_{yl}$) equal-sized data chunks denoted as $\{A_{l,i} \, | \, i \in [M_{xl}]\}$ ($\{B_{l,j} \, | \, j \in [M_{yl}]\}$)\footnote{These data chunks can themselves be thought of as tiles of the matrices.}. We can reverse the partitioning by concatenating the data chunks: $A_{\mathcal{S}_{Al}}^T = [A_{l,1}^T \, \ldots \, A_{l, M_{xl}}^T]$ and $B_{\mathcal{S}_{Bl}} = [B_{l,1} \, \ldots \, B_{l, M_{yl}}]$. Consider the $l$th computation layer. The corresponding task tile is of dimensions $|\mathcal{S}_{Al}| \times |\mathcal{S}_{Bl}|$. The above partitioning of $A_{\mathcal{S}_{Al}}$ and $B_{\mathcal{S}_{Bl}}$ also partitions the $l$th task tile into equally-sized (smaller) information tiles $A_{l,i} B_{l,j}$ each of dimensions $|\mathcal{S}_{Al}| / M_{xl} \times |\mathcal{S}_{Bl}| / M_{yl}$. We will apply coding (a polynomial code) across the data chunks $A_{l,i}$ and (separately) $B_{l,j}$ with the goal of recovering the information tiles. The information dimension of the code used in the $l$th layer will correspond to the number of the information tiles $\DM_l = M_{xl}M_{yl}$.

Two comments are in order. First, for the conceptual clarity for the moment we assume that $|\mathcal{S}_{Al}|$ and $|\mathcal{S}_{Bl}|$ are, respectively, much larger than $M_{xl}$ and $M_{yl}$, and so ignore integer effects. When we get to implementation we will need to deal with integer effects. Second, we choose $M_{xl}$ and $M_{yl}$ so that $a_l/(M_{xl}M_{yl})$ is (approximately) constant for all $l \in [\LY]$. While we need not make this choice, we make it to keep the quanta of computation (approximately) constant across layers. The implication is that information tiles will be of constant area. In particular, we choose there to be $\DM_{\text{sum}}=\sum_{l=1}^{\LY} \DM_l$ information tiles each of (approximate) area $\Aa\Bb/\DM_{\text{sum}}$. This assumption will prove useful when computing the response times of workers and when comparing to previous work. Note that the assumption that we keep $a_l/(M_{xl}M_{yl}) \approx \Aa\Bb/\DM_{\text{sum}}$ constant does {\em not} mean that the row- and column-dimensions of information tiles must be the same across different layers, only the area of each information tile is kept constant. This latter degree of flexibility will prove extremely useful in our overall design, especially when dealing with the integer constraints.

We now define the encoding functions. As previously mentioned we follow~\cite{YuEtAl:2017} and use polynomial codes. The polynomials used to encode the data chunks pertinent to the $l$th subtask are $ \hat{A}_l (x) = \sum_{i=1}^{M_{xl}} A_{l,i} x^{i-1}$ and $ \hat{B}_l (x) = \sum_{j=1}^{M_{yl}} B_{l,j} x^{(j-1)M_{xl}}$. For example, if $M_{xl} = 3$ and $M_{yl} = 4$ then $\hat{A}_l (x) = A_{l,1} + A_{l,2} x + A_{l,3} x^2$ and $\hat{B}_l (x) = B_{l,1} + B_{l,2} x^3 + B_{l,3} x^6 + B_{l,4} x^9$.


The $n$th worker receives $\LY$ pairs of encoded data chunks, $(\hat{A}_l(n),\hat{B}_l(n))$ for $ l \in [\LY]$.  The worker $n$ sequentially computes its $\LY$ jobs, $\hat{A}_1(n)\hat{B}_1(n)$ through $\hat{A}_{\LY}(n)\hat{B}_{\LY}(n)$, sending completed jobs to the master as soon as they are finished.

To recover all the information tiles that make up the $l$th layer of computation (and thus to recover the $l$th task tile), the master must receive at least $\DM_l = M_{xl}M_{yl}$ jobs from the $\ND$ workers, i.e., a subset of size at least $\DM_l$ of $\{\hat{A}_l(n)\hat{B}_l(n) \, | \, n \in [\ND]\}$.  We can conceive of each such small computational task as analogous to packet transmission through parallel and independent erasure channels where the code used in the $l$th channel is an $(\ND,\DM_l)$ MDS code. We use polynomial codes for the same reason as in previous work, namely that the computations of the polynomials provide the MDS property and only involve summation over data chuncks from $A$ (or $B$) and therefore is inexpensive when compared to the computation of the products $\hat{A}_l(n)\hat{B}_l(n)$.

One can notice that $\DM_1,\ldots \DM_{\LY}$ are design parameters that depend on the statistics of processing time. Given these parameters, and the shape of the information tiles in each layer, one then uses the procedure presented above to allocate tasks to workers. In the following sections we discuss how to chose the $\DM_l$ and the shape of the information tiles.

\section{Analytical evaluation}
\label{opt}

In this section, we assert a probabilistic model on the per-job finishing time distribution of individual workers. We then determine the finishing time distribution of our scheme as a function of the choice of $\{\DM_l\}_{l \in [\LY]}$. Then, we provide an approach to optimize the choice of the $\DM_l$. Lastly, we explain how these parameters are used in the design of a hierarchically coded solution to matrix multiplication. 


\subsection{Finishing time model}
\label{findopt}
The overall job of computing $AB$ is complete when each of the $\LY$ layers completes. For layer $l$ to complete, at least $\DM_l$ workers must finish their $l$th  task. In the following we determine the distribution of (at least) this minimal number of tasks completing each layer. This is the \emph{finishing time}. 

In our analysis we assume that workers complete tasks according to the shifted exponential distribution, previously used in~\cite{Lee:TIT17}. Let $T_1,\ldots, T_{\ND}$ be independent and identical distributed shifted exponential random variables with scale parameter $\mu$ and shift parameter $\alpha$. $T_{n}$ denotes the time the $n$th worker takes to compute $AB$ product on its own. Thus, the probability that a worker is able to finish the $AB$ product by time $t$ is $\mathbb P(T_{n}<t) = 1-e^{-\frac{1}{\mu}\left(t-\alpha\right)}$ for $t\geq \alpha$, and $\mathbb P(T_{n}<t) = 0$ else. We subdivide $AB$ into $\DM_{\text{sum}}$ equal-sized information tiles. For each layer each worker must compute the equivalent of one (encoded) information tile. The model of computation is conditionally deterministic: the $n$th worker completes one task every $T_n/\DM_{\text{sum}}$ seconds. The realization of the $T_n$ sets the speed of the workers. Once those speeds are set each worker is assumed to process equally-sized jobs in a (conditionally) deterministic fashion. Therefore, for the $n$th worker to finish tasks through the $l$th layer takes $lT_n / \DM_{\text{sum}}$ seconds. Let $T_{n:\ND}$ be the $n$th order statistics and $\tau$ denote the finishing time. Then the expected finishing time is, 
\begin{align}\label{eqn:avg.finish}
	\mathbb E[\tau] &= \max_{l\in[\LY]} \frac{l}{\DM_{\text{sum}}} \mathbb E[T_{\DM_l:\ND}] \nonumber \\
	& \approx  \max_{l\in[\LY]} \frac{l}{\DM_{\text{sum}}}\left( \alpha + \mu \log \left(\frac{\ND}{\ND-\DM_l}\right)\right).
\end{align}

In the extended version of this paper, we will provide the detailed proof to (\ref{eqn:avg.finish}), and will discuss how to optimize the choice of $\{\DM_l \}_{l\in[\LY]}$ to minimize the expected finishing time.

  


\subsection{Practical implementation}
Besides $\DM_l$, we can further optimize the $M_{xl}$ and $M_{yl}$ parameters to reduce the amount of data that the master needs to distribute to the workers. While, due to space constraints, we leave this discussion to future work, we now present the algorithm used to minimize the communications. Algorithm~\ref{Alg} selects tiles by iteratively placing  rectangles of area $\floor{\frac{\Bb}{\DM_{\text{sum}}}}\DM_l \Aa$ in an $\Aa \times \Bb$ rectangle. The algorithm avoids creating overlaps with previously placed rectangles. The rectangle in the $l$th iteration is then partitioned into $\DM_l$ equally size tiles.

\begin{algorithm}
\caption{Partitioning of a $\Aa\times \Bb$ rectangle
  into $\LY$ task tiles given  profile $\{\DM_1,\ldots,
  \DM_{\LY} \}$, where $\DM_{l-1} \geq \DM_l$, $\DM_{\text{sum}} = \sum_{l=1}^\LY \DM_l$, and $\Aa,\AB,\Bb \gg \DM_l$ for all $l \in [\LY]$. Partition the $l$th task tile into information tiles given $\{M_{xl},M_{yl}\}$.}\label{Alg}
\Input{$\LY, \{\DM_l,M_{xl},M_{yl}\}_{l \in [\LY]}, \Aa, \AB, \Bb, \DM_{\text{sum}}$}
\label{Alg}
\begin{algorithmic}[1]
\BState \textbf{for} $l \in [\LY]$:
\State \multiline {Slice the $l$th task tile from the remaining, un-allocated, rows and columns such that the $l$th task tile contains all $\Aa$ rows and the next leftmost $\floor{\frac{\Bb}{\DM_{\text{sum}}}}\DM_l$ columns.}
\State \multiline {Given $\{M_{xl},M_{yl}\}$, decompose the $l$th task tile into $\DM_l$ equally sized information tiles.}
\State \textbf{end for}
\end{algorithmic}
\end{algorithm}


We comment that the rounding error in Alg.~\ref{Alg} results in $\Bb-\sum_l ( \frac{\Bb}{\DM_{\text{sum}}}-1 )\DM_l=\DM_{\text{sum}}$ extra columns.  These requires $\Aa\AB\DM_{\text{sum}}$ additional computation to multiply $A$ and the $\DM_{\text{sum}}$ last columns of $B$. We assign this negligible computation (negligible since $\Aa\Bb\DM_{\text{sum}} \ll \Aa\AB\Bb$) to the master. 

\section{Evaluation}
\label{sim}
We now numerically evaluate the performance of our scheme. We first
consider the shifted exponential model of Sec.~\ref{opt}. We then
evaluate the performance on Amazon EC2. We compare our results with
polynomial~\cite{YuEtAl:2017} and sum-rate~\cite{EXPLOIT:ISIT18}
codes.

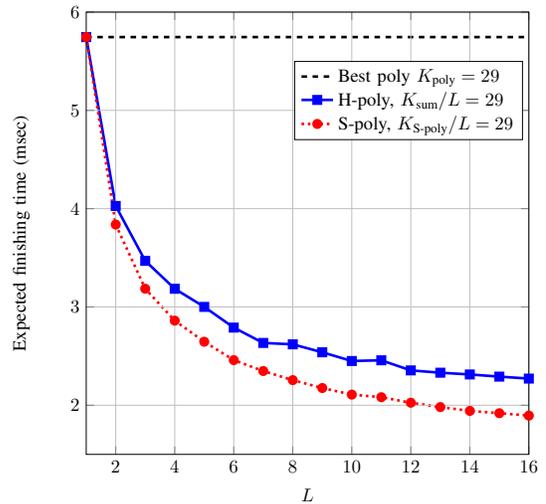
\begin{figure}
\centering
		\begin{tikzpicture}[scale=0.699]
	\begin{axis}[
	height=10cm,
	width=10cm,
	grid=major,
	xlabel={$L$},
	ylabel={Expected finishing time (msec)},
legend style={at={(0.47,0.8)},anchor=west,nodes=right},	
	axis on top,xmin=1, xmax=16, ymin=1.5, ymax=6]
	
	\addlegendentry{Best poly $\DM_{\text{poly}}=29$}
	\addplot [line width=0.5mm, color=black, dashed, every mark/.append style={solid, fill=black}] coordinates {
 	 	(1,5.74561  )
 	(2,5.74561  )
 	(3,5.74561  )
 	(4,5.74561  )
 	(5,5.74561  )
 	(6,5.74561)
 	(7,5.74561  )
 	(8,5.74561  )
 	(9,5.74561  )
 	(10,5.74561  )
 	(11,5.74561  ) 
 	(12,5.74561)
 	(13,5.74561  )
 	(14,5.74561  )
 	(15,5.74561  )
 	(16,5.74561)
 	};
 	
	\addlegendentry{H-poly, $\DM_{\text{sum}}/\LY=29$}
		\addplot [line width=0.5mm, color=blue, solid, every mark/.append style={solid, fill=blue},mark=square*] coordinates {
     	(1,5.74561  )
 	(2,4.0275   )
 	(3,3.46891  )
 	(4,3.18492  )
 	(5,3.00061  )
 	(6,2.78975)
 	(7,2.63376  )
 	(8,2.61969  )
 	(9,2.53872  )
 	(10,2.44999  )
 	(11,2.4572   ) 
 	(12,2.35434)
 	(13,2.33095  )
 	(14,2.31307  )
 	(15,2.29103  )
 	(16,2.27112) 
	}; 	
		\addlegendentry{S-poly, $\DM_{\text{S-poly}}/\LY=29$}
		\addplot [line width=0.5mm, color=red, dotted, every mark/.append style={solid, fill=red},mark=otimes*] coordinates {
 	(1,5.74561)
 	(2,3.83812)
 	(3,3.18557)
 	(4,2.86076)
 	(5,2.64613)
 	(6,2.45903)
 	(7,2.34816)
 	(8,2.25542)
 	(9,2.17506)
 	(10,2.10813)
 	(11,2.0817) 
 	(12,2.02558)
 	(13,1.98122)
 	(14,1.94242)
 	(15,1.91907)
 	(16,1.8946)     
	}; 	
 	
	\end{axis}
	\end{tikzpicture} 
	\setlength{\belowcaptionskip}{-12pt}
\caption{The expected finishing time vs. number of layers for the shifted exponential distribution, where $\ND = 200$ and $(\mu,\alpha) = (1,0.01)$.}
\label{fin_vs_L}
\end{figure}

Fig.~\ref{fin_vs_L} plots the expected finishing time vs. number of
layers $\LY$ based on the shifted exponential distribution model with
$\mu=1$ and $\alpha=0.01$. In each trial we generate $\ND$ independent exponential random variable $T_n$, $n \in [\ND]$, one per worker, each according to scale parameter $\mu > 0$ and shift parameter $\alpha$. The best\footnote{The best code is the code that has the
  minimum $\mathbb E[\tau]$ for the parameters given in the
  caption (minimized over different choices of $\DM_{\text{poly}}$).} polynomial
code, corresponding to $\DM_{\text{poly}}=29$, achieves an expected
finishing time of $5.7$ msec. For the same per-worker
computation load ($\DM_{\text{sum}}/\LY=29$), we plot (the solid line) the
performance of the hierarchical code for different choices of
$\LY$. The decrease in $L$ illustrates the fact that division of the
job into smaller information tiles (larger $\LY$) results in an
acceleration of the computation of $AB$.  This observation will hold
as long as the profile $\{\DM_l \}_{l \in [\LY]}$ is set
appropriately. In particular, we observe a $60\%$ improvement in
expected finishing time for $\LY=16$ when compared to the best
polynomial codes. Finally, we plot (the dotted line) the expected
finishing time of sum-rate coding for different $\LY$. It can be
observed that the performance of sum-rate coding lower bounds
hierarchical coding. The gap between hierarchical codes and this bound
increases as $\LY$ increases. This can be explained as follows. If the
number of layers is smaller (e.g., $\LY=2$) we have fewer options to
set the profile of the hierarchical code and, with higher probability,
we select the optimal profile that is close to the pattern of
completions realized by the sum-rate code. However, as $\LY$
increases, the flexibility to select optimal profile increases and
chance to getting close to the optimal pattern decreases.

We note, however, that $\LY$ is a design parameter. In practice,
excessively increasing the number of layers $\LY$ is not advisable due
to the increase in decoding complexity. Decoding complexity is
governed by the complexity of interpolating a degree-$k$ polynomial,
which is order $\mathcal{O}(k \log^2 k)$~\cite{YuEtAl:2017}. 

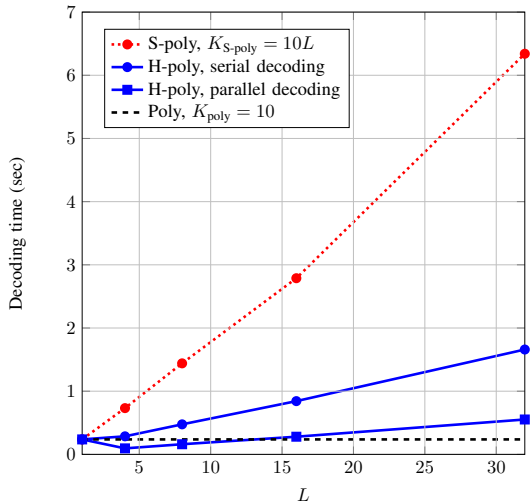
\begin{figure}
\centering
		\begin{tikzpicture}[scale=0.699]
	\begin{axis}[
	height=10cm,
	width=10cm,
	grid=major,
	xlabel={$L$},
	ylabel={Decoding time (sec)},
legend style={at={(0.05,0.85)},anchor=west,nodes=right},	
	axis on top,xmin=1, xmax=32, ymin=0, ymax=7.000000000]

		\addlegendentry{S-poly, $\DM_{\text{S-poly}} =10\LY$}
		\addplot [line width=0.5mm, color=red, dotted, every mark/.append style={solid, fill=red},mark=otimes*] coordinates {
 	(1, 0.236892104)
 	(4, 0.732095351)
 	(8, 1.440659991)
 	(16,2.788075078) 
 	(32,6.340680892) 	   
	};

	\addlegendentry{H-poly, serial decoding}
		\addplot [line width=0.5mm, color=blue, solid, every mark/.append style={solid, fill=blue},mark=*] coordinates {
 	(1, 0.236892104)
 	(4, 0.2857)
 	(8, 0.4753)
 	(16,0.8428)
 	(32,1.6596) 	
	};

	
	\addlegendentry{H-poly, parallel decoding}
		\addplot [line width=0.5mm, color=blue, solid, every mark/.append style={solid, fill=blue},mark=square*] coordinates {
 	(1, 0.236892104)
 	(4, 0.0951)
 	(8, 0.1605)
 	(16,0.2795)
 	(32,0.5522)
	};	
	
	\addlegendentry{Poly, $\DM_{\text{poly}}=10$}
	\addplot [line width=0.5mm, color=black, dashed, every mark/.append style={solid, fill=black}] coordinates {
 	(1, 0.236892104)
 	(4, 0.236892104)
 	(8, 0.236892104)
 	(16,0.236892104)
 	(32,0.236892104)
 	}; 	
	

	\end{axis}
	\end{tikzpicture} 
	\setlength{\belowcaptionskip}{-12pt}
\caption{The decoding time vs number of layers in Amazon EC2 where $(\Aa,\AB,\Bb)=(1000,1000,1000)$ and $\DM_{\text{sum}}=10\LY$.}
\label{dec_vs_L}
\end{figure}

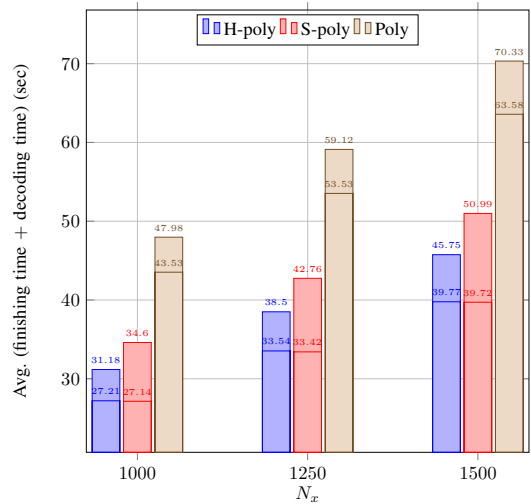
\begin{figure}
\centering
	\begin{tikzpicture}[scale=0.699]
\begin{axis}[
	height=10cm,
	width=10cm,
	grid=major,
    ybar,
    bar width=15,
    enlargelimits=0.15,
    legend style={at={(0.5,0.99)},
      anchor=north,legend columns=-1},
    ylabel={Avg. (finishing time $+$ decoding time) (sec)},
    xlabel={$\Aa$},
    symbolic x coords={$1000$,$1250$,$1500$},
    every node near coord/.append style={font=\tiny},
    xtick=data,
    nodes near coords,
    nodes near coords align={vertical},
    ]

\addplot coordinates {($1000$,   27.207)($1250$,33.535) ($1500$,39.774 )($1000$,   31.177)($1250$,38.502) ($1500$,45.753)};
\addplot coordinates {($1000$,   27.144)($1250$,33.415) ($1500$,39.719) ($1000$,   34.604)($1250$,42.755) ($1500$,50.989 )};
\addplot coordinates {($1000$,   43.529)($1250$,53.531) ($1500$,63.580)($1000$,   47.975)($1250$,59.117) ($1500$,70.331)};
\legend{H-poly,S-poly,Poly}
\end{axis}
\end{tikzpicture}
	\setlength{\belowcaptionskip}{-12pt}
\caption{The sum of average finishing (computation) and decoding times of matrix multiplication of different dimensions. The lower bar is the average finishing time; the upper is sum of average finishing and decoding times; and the difference is decoding time.}
\label{EC2} 
\end{figure}

We now discuss the Amazon EC2 results, presented in
Figs.~\ref{dec_vs_L} and~\ref{EC2}.  We implemented decoding on a
``t2.micro'' instance.  The decoder solves a system of linear
equations which involves a Vandermonde matrix. Both $A$ and $B$ are
$1000 \times 1000$ matrices and the average recovery threshold per
layer (respectively, $\DM_{\text{poly}}, \DM_{\text{S-poly}}/\LY,
\DM_{\text{sum}}/\LY$ for the three schemes) is set to $10$. In Fig.~\ref{dec_vs_L}
we plot decoding time versus number of layers. We plot (solid lines)
the serial and parallel decoding times when using hierarchical codes. Each data point on these lines corresponds to different number of layers, $\LY \in \{1,4,8,16,32\}$. The profile of hierarchical coding when using a single layer ($\LY=1$) is $\DM_1=10$. For $\LY > 3$, this profile is set as follows. The recovery threshold of the first three layers are set to be the largest, i.e., $\DM_{i}=3\LY+1$ for $i \in [3]$, and $\DM_{j}=1$ for the remaining layers ($j \in [\LY-3]+3$). One can observe that hierarchical coding can achieve a small decoding time close to that of polynomial codes when the decoding of each layer is conducted in parallel. While the decoding time of hierarchical coding when decoding is carried out serially is larger than that of polynomial codes, the decoding time of sum-rate codes is the largest and increases dramatically as $\LY$ is increased. 

In Fig.~\ref{EC2}, we plot the sum of average finishing (computation) and decoding times. For example, in the left hand plot it takes on average $27.21$ sec for hierarchical polynomial code to finish enough per-layer computations and $31.18$ sec to recover the matrix multiplication; where the average time to decode is $3.97$ sec. This experiment is  performed in Amazon EC2 with $\LY=2, \ND=16,$ and $\DM_{\text{poly}}= \DM_{\text{S-poly}}/L=\DM_{\text{sum}}/L=10$. We used $\ND+1$ t2.micro instances and implemented our approach in C++ using (the relatively slow) Eigen library to perform matrix multiplication. Since in EC2 we rarely observe stragglers in small-scale distributed system (our system includes 16 workers and the master), we artificially delay nodes so that our design can be tested. In this ``artificial-straggler'' scenario we assign workers to be stragglers independently with probability
$0.5$. Workers that are designated stragglers are assigned one more extra computation than non-stragglers per layer (i.e., stragglers are two times slower than non-stragglers). We repeated this experiment for $10$ iterations and in each iteration we
log the computation time required by each node to complete its
subtasks. We then add the decoding time at the master. For three different matrix multiplication problems, we
measure the sum of average finishing and decoding times of three distinct schemes: polynomial coding, hierarchical
polynomial coding with profile $(14,6)$, and sum-rate
polynomial coding. For each of the three problems we use matrix dimensions $(\AB,\Bb)=(50,20000)$ and respectively $\Aa$ dimensions $1000,1250,$ and $1500$. The average finishing times of sum-rate and hierarchical coding are approximately the same, while the sum of average finishing and decoding times of the hierarchical approach yields a speedup of $5.24$ sec when compared to sum-rate approach for $\Aa=1500$.
The hierarchical coding achieves $35\%$ improvements in comparison to polynomial coding.

\section{Conclusion}
In this paper we introduce hierarchical coded matrix multiplication. Through this hierarchical design, our scheme can exploit the work completed by all workers, including stragglers. To apply hierarchical coding into matrix multiplication, we connect the task allocation problem that underlies coded matrix multiplication to a rectangle partitioning problem. In Amazon EC2, our scheme achieves $35\%$ improvement
in the average finishing time when compared to the previous approach.

\bibliographystyle{IEEEtran} 
\bibliography{reference}
\end{document}